\documentclass[twocolumn,showpacs,preprintnumbers,prl,amsmath,amssymb,superscriptaddress]{revtex4}

\usepackage{graphicx}
\usepackage{dcolumn}
\usepackage{bm}

\begin{document}


\title{Coherent Control of Ultracold Molecule Dynamics in a\\
Magneto-Optical Trap Using Chirped Femtosecond Laser Pulses}

\author{Benjamin L. Brown}
 \email{ben.brown@nist.gov}
 \affiliation{Clarendon Laboratory, Department of Physics, University of Oxford, Oxford, OX1 3PU, United Kingdom}
 \affiliation{The Institute of Optics, University of Rochester, Rochester, New York 14627}

\author{Alexander J. Dicks}
 \affiliation{Clarendon Laboratory, Department of Physics, University of Oxford, Oxford, OX1 3PU, United Kingdom}

\author{Ian A. Walmsley}
 \affiliation{Clarendon Laboratory, Department of Physics, University of Oxford, Oxford, OX1 3PU, United Kingdom}

\date{\today}

\begin{abstract}
We have studied the effects of chirped femtosecond laser pulses on
the formation of ultracold molecules in a Rb magneto-optical trap.
We have found that application of chirped femtosecond pulses
suppressed the formation of $^{85}\mathrm{Rb}_2$ and
$^{87}\mathrm{Rb}_2$ $a^3\Sigma_u^+$ molecules in contrast to
comparable non-chirped pulses, cw illumination, and background
formation rates.  Variation of the amount of chirp indicated that
this suppression is coherent in nature, suggesting that coherent
control is likely to be useful for manipulating the dynamics of
ultracold quantum molecular gases.
\end{abstract}


\maketitle


Achieving control of the dynamics of quantum systems has been a
long-standing goal of physics and
chemistry~\cite{WARREN1993,RABITZ2000}. Rapid advances in the
manipulation of laser-matter interactions to obtain desired outcomes
by means of tailored optical fields have been enabled by the
development of ultrafast femtosecond optical sources and
pulse-shaping techniques. In particular, recent successes include
exciting \textit{a priori} specified quantum states in
atoms~\cite{WEINACHT1999} and molecules~\cite{BARDEEN1997}, as well
as in selectively cleaving chemical bonds in complex
molecules~\cite{ASSION1998}. Simultaneously, a very different thrust
has been extending  the regime of the ultracold ($T \leq
1\,\mathrm{mK}$) to simple molecular complexes~\cite{BURNETT2002}.
Robust samples of trapped ultracold molecules are expected to
facilitate significant advances in molecular spectroscopy, collision
studies, and perhaps quantum
computation~\cite{DEMILLE2002,TESCH2002}.

The difficulty in generalizing laser cooling techniques to molecules
has stimulated exploration of alternative approaches to producing
ultracold molecules.  One approach, using magnetic Feshbach
resonances, has led to the observation of molecular Bose-Einstein
condensates~\cite{JOCHIM2003,GREINER2003,ZWIERLEIN2003}.  The study
of other routes to ultracold molecule formation has also been a
topic  of intense activity~\cite{MASNOUSEEUWS2001}.  An extremely
successful optical approach has been to photoassociate molecules
from ultracold atoms~\cite{THORSHEIM1987}. In this method, a sample
of ultracold atoms is irradiated by a cw laser tuned to excite free
atoms to weakly bound excited states. Stable ground-state molecules
may then form by spontaneous emission, provided the Franck-Condon
overlap factors are favorable. Ultracold molecules have been
observed using this technique for a variety of
homonuclear~\cite{FIORETTI1998,NIKOLOV1999,GABBANINI2000,FATEMI2002}
and
heteronuclear~\cite{KERMAN2004,MANCINI2004,HAIMBERGER2004,WANG2004A}
alkali metal species.

\begin{figure}[b]
\includegraphics{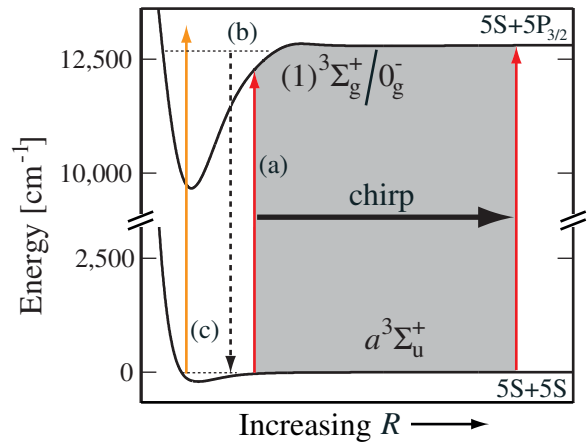}
\caption{\label{fig:potentials} (Color online) Schematic of the
chirped femtosecond pulse photoassociation experiment.  The
potential energy curves of $\mbox{Rb}_2$ involved in the design of
the experiment are shown. (a) Ultracold atoms are excited over a
range of internuclear separations by a chirped femtosecond pulse to
bound states below the $5\mathrm{S}$+$5\mathrm{P}_{3/2}$
dissociation limit.  The time-dependent frequency sweep ensures the
unlikelihood of population cycling back to the ground state.  (b)
Excited molecules undergo spontaneous decay, some forming stable
ground-state molecules.  (c) These molecules are detected by
resonance-enhanced two-photon ionization and time-of-flight ion-mass
spectroscopy.}
\end{figure}

Recent proposals have suggested controlling the interactions between
ultracold atoms with tailored picosecond optical
fields~\cite{VALA2001A,LUCKOENIG2004A,LUCKOENIG2004,KOCH2004}.
The possibility of improved formation of ultracold molecules using
femtosecond coherent control techniques is therefore enticing.
When a pulse length is $< 100\,\mathrm{ps}$, the light-matter
interaction is entirely coherent, and therefore the details of the
temporal shape of the pulse electric field may effect the
interaction significantly.
This Letter describes the first experiments designed to enhance the
molecule formation process in a magneto-optical trap
(MOT)~\cite{RAAB1987} with chirped femtosecond pulses.
%
%
A schematic of the experiments is shown in
Fig.~\ref{fig:potentials}.
The broad bandwidth of femtosecond pulses permits excitation over a
longer range of internuclear distances than picosecond pulses, thus
addressing a larger number of atom pairs.
%
%
Introducing a positively chirped molecular ``$\pi$''-pulse allows
wave packet dynamics to strongly suppress detrimental population
cycling~\cite{CAO1998}.
This method also provides a route to optimal control of ultracold
molecule formation~\cite{BROWN2005}.
Recent experiments using picosecond~\cite{FATEMI2001} and chirped
nanosecond~\cite{WRIGHT2005} pulses for photoassociation (PA)
provide encouragement for this approach.
However, as we show in this Letter, the dynamics of molecule
formation using femtosecond pulses appears to be significantly
different from those predicted for picosecond
pulses~\cite{LUCKOENIG2004A,LUCKOENIG2004}.  Here we report a
suppression of molecule formation rather than an enhancement;
nevertheless, the mechanism responsible for this effect appears to
be coherent.


Femtosecond (fs) chirped-pulse PA was undertaken in an Rb MOT
because its trapping transition ($780.27\,\mathrm{nm}$) is near the
peak of the gain profile for Ti:sapphire ($800\,\mathrm{nm}$). Each
isotope of Rb could be trapped by adjustment of the MOT trapping and
repumping laser frequencies.
Typical traps contained $2 \times 10^{7}$ atoms in a spheroidal
cloud of diameter $0.7\,\mathrm{mm}$, yielding a peak density on the
order of $10^{10}\,\mbox{cm}^{-3}$.  The temperature of the atoms in
the MOT was measured to be $170\,\mu\mathrm{K}$ using the release
and recapture method~\cite{LETT1988}.

Two different PA lasers were employed to study the effects of light
on the molecule formation process in the MOT. The first was a cw
free-running diode laser (Sanyo DL7140-201) with an output power of
$56\,\mathrm{mW}$, temperature-tuned between
$700$--$785\,\mathrm{nm}$.  The second laser was a commercial
femtosecond-pulse oscillator (Spectra-Physics Mai Tai) whose carrier
wavelength was tunable between $750$--$850\,\mathrm{nm}$ with output
power $\leq 800\,\mathrm{mW}$ over this range. This laser had a
repetition rate of $80\,\mathrm{MHz}$ and produced Gaussian-envelope
transform-limited pulses of temporal intensity full-width at
half-maximum (FWHM) $<100\,\mathrm{fs}$, corresponding to a spectral
bandwidth FWHM $> 150\,\mathrm{cm}^{-1}$ for the available carrier
wavelength range. Its output spectrum was monitored with a
spectrometer and its field profile was regularly characterized using
Spectral Interferometry for Direct Electric Field Reconstruction
(SPIDER)~\cite{IACONIS1998}.


Molecules formed in the MOT were detected via time-of-flight (TOF)
ion-mass spectroscopy using a channel electron multiplier.  A
narrow-bandwidth ($< 3\,\mathrm{GHz}$) tunable pulsed dye laser
($9\,\mathrm{ns}$ pulse duration, $50\,\mathrm{Hz}$ repetition rate)
ionized the MOT cloud with pulses of energy $500\,\mu\mathrm{J}$.
The PA and ionization lasers were merged onto the same beampath with
a dichroic beamsplitter and focused at the MOT cloud, ensuring that
the sample that we ionized was the same as that affected by our PA
lasers.

We chose to detect ground triplet state $a^{3}\Sigma_{u}^{+}$
$\mbox{Rb}_{2}$ molecules, which have been observed to form
spontaneously in the MOT through one or both of two channels: PA
stimulated by the MOT lasers themselves, and three-body
recombination~\cite{GABBANINI2000,CAIRES2005}.  The ionization
wavelength was tuned to $602.7\,\mathrm{nm}$, which ionizes ground
triplet state molecules via a resonant two-photon excitation
($X^{2}\Sigma_{g} \leftarrow (2)^{3}\Pi_{g} \leftarrow
a^{3}\Sigma_{u}^{+}$).  To ensure that the detected molecules were
in their electronic ground state, the MOT and PA lasers were shut
off for at least $1\,\mu\mathrm{s}$ before the arrival of the
ionization laser pulse to allow the atoms and molecules in the MOT
to decay to their ground states.
%
%
For a $^{85}\mbox{Rb}$ MOT the background detection rate of
$\mbox{Rb}^{+}_{2}$ ions was typically 2--3 ions per pulse, while
for a $^{87}\mbox{Rb}$ MOT the rate was significantly lower
(0.3--0.8 ions per pulse).


A typical experiment involved applying one of the PA lasers to the
MOT, applying the ionizing laser, and accumulating ion counts for a
set number of ionization laser shots $N_{\mathrm{shots}}$.
A background scan with the PA laser blocked was then immediately
acquired for another $N_{\mathrm{shots}}$ to determine the relative
effect of the PA laser.


The cw PA laser was focused on the MOT with an intensity of around
$10^3\,\mathrm{W}\,\mathrm{cm}^{-2}$.  For an $^{87}\mbox{Rb}$ MOT,
the molecular signal was increased for detunings of a few
$\mbox{cm}^{-1}$ below the trapping transition, with significant
enhancement (by a factor of 1.5--2) at frequencies corresponding to
resonant transitions to excited molecular vibrational states. For
$^{85}\mbox{Rb}$, the molecular signal was unaffected at resonant
transition frequencies, and generally reduced below the background
level by 50\% at non-resonant frequencies.  These results confirm
those of Ref.~\cite{GABBANINI2000}, where the quenching phenomenon
was interpreted as a coupling of molecules that form spontaneously
in the $^{85}\mbox{Rb}$ MOT to excited dissociative states.



In order to promote efficient photoassociative excitations near the
D2 line by the fs pulses, care was taken to select a pulse spectrum
with power concentrated in the region just below the
$5\mathrm{S}$+$5\mathrm{P}_{3/2}$ dissociation limit
($780\,\mathrm{nm}$). The carrier wavelength of the fs laser was
tuned to $783\,\mathrm{nm}$.  A
spectral filter was used to ensure that the broad spectrum of the fs
pulses did not contain significant power on the blue side of the D2
line (which would lead to direct excitation of atoms to repulsive
states lying above the $5\mathrm{S}$+$5\mathrm{P}_{3/2}$
dissociation limit).
The fs pulse power was attenuated to $\leq 0.3\,\mathrm{nJ}$ per
pulse, corresponding to focused peak intensities on the order of
$10^7\,\mathrm{W}\,\mathrm{cm}^{-2}$.
Since this intensity is many orders of magnitude above the Rb D2
atomic transition saturation intensity
($6\,\mathrm{mW}\,\mathrm{cm}^{-2}$), the light-matter interaction
is beyond the perturbative regime.
%

\begin{figure}
\includegraphics{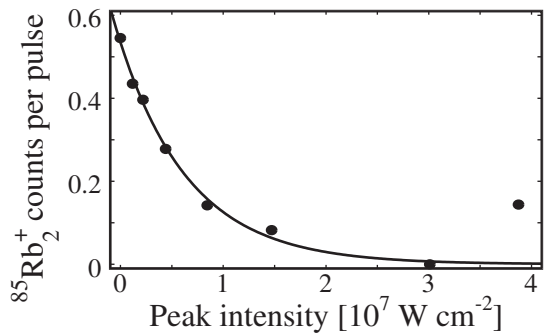}
\caption{\label{fig:NDscan} $^{85}\mathrm{Rb}^{+}_{2}$ counts per
ionization pulse vs.\ peak intensity of the applied fs pulses
(filled circles). Each data point represents the accumulated result
of an identical number of ionization laser shots. An exponential
decay fit of the data is shown (solid line).}
\end{figure}

Application of the spectrally-filtered fs-pulse PA laser to
$^{85}\mbox{Rb}$ and $^{87}\mbox{Rb}$ MOTs generally caused a strong
reduction in the molecular ion signal as compared to background. In
the case of a $^{85}\mbox{Rb}$ MOT, the quenching observed due to
the fs PA laser was stronger than that induced by the cw PA laser.
In the case of a $^{87}\mbox{Rb}$ MOT, contrary to the augmentative
effect observed when the cw PA laser was applied, $\sim 50\%$ fewer
molecular ions with respect to background were observed upon
application of the fs laser under similar experimental conditions.
$^{85}\mathrm{Rb}_2^+$ ionization spectra for
600--610$\,\mathrm{nm}$ recorded for the cases of no PA laser, cw PA
laser, and fs PA laser had similar structure.
Variation of the average power of the fs PA laser revealed evidence
of an exponential-decay dependence of molecular ion yield on pulse
power (see Fig.~\ref{fig:NDscan}).


We further explored the effects of fs pulses on the molecular
formation rate by varying the spectral chirp of the pulses.  The
$100\,\mathrm{fs}$ FWHM pulses were chirped by passing the beam 13
times through a $5.0\,\mathrm{in.}$-length plane-parallel dispersive
glass block.
The chirped pulses had an estimated temporal intensity FWHM of
$5.8\,\mathrm{ps}$.
To test whether the quenching effect involved resonant excitation of
ground state atoms by the fs laser, we also tuned its carrier
wavelength to $850\,\mathrm{nm}$ (as far to the red as
experimentally feasible) and removed the spectral filter.  A
spectrometer was used to confirm that the power spectrum of the
chirped pulses did not differ substantially from that of non-chirped
pulses.

\begin{figure}
\includegraphics{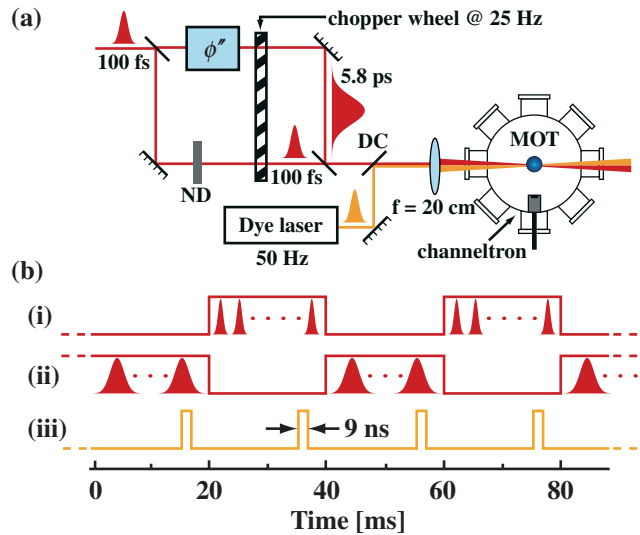}
\caption{\label{fig:setup}(Color online) (a) Schematic of the
experimental setup. The fs PA beam was split by a 50:50
beamsplitter; one beam was chirped by the dispersive glass block
($\phi''$), and the other (reference) beam was power-balanced using
neutral density filters (ND).  A chopper wheel synchronized with the
detection apparatus allowed only one of the two beams to pass at a
given instant.  The two beams were merged at a second 50:50
beamsplitter and focused to the MOT. The dye laser was merged onto
the same beampath with a dichroic beamsplitter (DC).  (b)
Reference/chirp beam detection timing. Alternating
$80\,\mathrm{MHz}$ trains of reference (b-i) and chirped (b-ii)
pulses were applied to the MOT. One dye laser ionization pulse
(b-iii) fired for each train of pulses.  Not shown here is the
extinction of each pulse train for $1\,\mu\mathrm{s}$ prior to the
firing of the dye pulse.}
\end{figure}

Due to run-to-run drift of the MOT conditions and the significant
time required to re-align the setup when removing and re-inserting
the glass block into the PA laser beampath, we implemented a
``real-time'' differential measurement to compare the effects of
chirped and non-chirped fs pulses.  A schematic diagram of the
experimental setup appears in Fig.~\ref{fig:setup}. A broadband
50:50 beamsplitter was used to split the fs laser beam into two
identical components. The dispersive glass block was inserted into
the path of one of these beams (hereafter the ``chirped beam''). The
second, transform-limited, beam (the ``reference beam'') was
attenuated with absorptive neutral density filters so that each beam
had the same average power, to $\pm1\%$ accuracy.  An optical
chopper wheel was aligned such that at any given instant one beam
was blocked and the other was allowed to pass. Downstream from the
chopper, the two beams were merged onto the same beampath using a
second 50:50 broadband beamsplitter. From this point forward, the
two beams were aligned and focused to the MOT position as usual.

The experiment timing was set up so that for a single rotation of
the chopper wheel, one chirped pulse experiment and one reference
pulse experiment occurred in sequence (see Fig.~\ref{fig:setup}).
The TOF data acquisition was performed in ``toggle'' mode: TOF
traces acquired for chirped pulse cases were added to the TOF
histogram and traces acquired for reference pulse cases were
subtracted from the histogram, yielding a net difference TOF
histogram.
A baseline differential measurement for the case with the dispersive
glass block removed in Fig.~\ref{fig:setup} revealed a difference
between two identical beams of fewer than $0.04\,\mbox{Rb}_{2}^{+}$
ions detected per ionization pulse.

\begin{figure}
\includegraphics{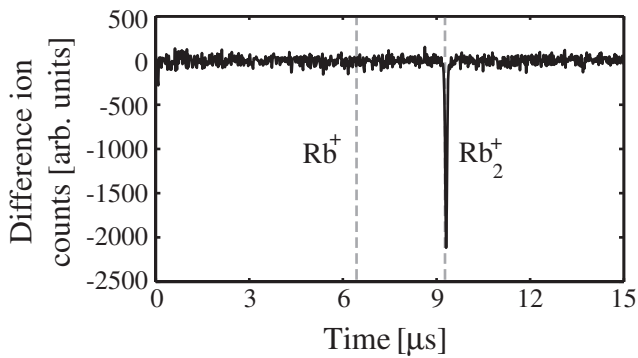}
\caption{\label{fig:toggledata}  Difference (``chirped case'' $-$
``reference case'') TOF histogram comparing the effects of chirped
and transform-limited reference pulses.  The vertical dashed lines
indicate the approximate arrival times of $^{85}\mathrm{Rb}$ atomic
and molecular ions at the channeltron. There is no difference in the
atomic ion signal between the two cases.  A significant negative
feature appears at the molecular ion arrival time, indicating that
more $\mbox{Rb}_{2}^{+}$ ions were detected after applying the
reference pulses than the chirped pulses.}
\end{figure}

The results of an experiment comparing the effects of a pulse train
of $5.8\,\mathrm{ps}$ pulses to the effects of a pulse train of
$100\,\mathrm{fs}$ pulses are shown in Fig.~\ref{fig:toggledata}.
The differential TOF trace reveals no discernable difference in the
number of atomic ions, and a sharp negative feature in the vicinity
of the arrival time of $^{85}\mbox{Rb}_{2}^{+}$ ions.  The magnitude
of this feature exceeded the baseline experiment $\mbox{Rb}_{2}^{+}$
detection rate by~400\%. The negative feature in
Fig.~\ref{fig:toggledata} signifies that appreciably more
$\mbox{Rb}_{2}^{+}$ ions were accumulated after application of the
transform-limited reference pulses.
Thus, the chirped pulses quenched the $\mbox{Rb}_{2}^{+}$ ion signal
more than the reference pulses. Since the chirped pulses had a peak
intensity $\sim70$ times smaller than the reference pulses, this
quenching effect was not solely dependent on the peak intensity, but
also on the phase of the applied optical field.
The quenching tended to increase with increasing chirp magnitude,
although this trend was not monotonic as some chirp values were
significantly more effective than others. These results provide
evidence that the quenching effect is a coherent phenomenon in which
wave packet dynamics play some role.

The exponential decay dependence of molecular formation rate on
applied fs pulse peak intensity is consistent with a single-photon
excitation loss from either the detected $a^3\Sigma_u^+$ or the
excited $(1)^3\Sigma_g^+\big/0_g^-$ states.
There are several coherent excitation pathways for
$850\,\mathrm{nm}$ light that could be responsible for a reduction
in the number of detected $a^3\Sigma_u^+$ molecules.  One is
MOT-laser PA followed by excitation of excited molecules in the
$(1)^3\Sigma_g^+\big/0_g^-$ states to bound states in potentials
that dissociate to 5P+5P ($(6)^3\Sigma_u^+$, $(7)^3\Sigma_u^+$, and
$(4)^3\Pi_u$)---all of which are ``dark'' to our detection
scheme---or to dissociative states lying above the 5S+6P
dissociation limit.  Another possibility is that a wave packet is
excited from the ground state $a^3\Sigma_u^+$ onto the $C_3R^{-3}$
potential of the $(1)^3\Sigma_g^+$ state.
This wave packet may have enhanced Franck-Condon overlaps with
either continuum states lying above the 5S+5S dissociation limit or
undetected $a^3\Sigma_u^+$ bound states, thus resulting in a
decrease in the number of detected $a^3\Sigma_u^+$
molecules~\cite{KOCH2005A}.


In summary, we observed that application of fs pulses designed to
photoassociate ultracold atoms results in a decrease, rather than an
increase, in the formation of stable ultracold $a^3\Sigma_u^+$
molecules.
Related work has recently been undertaken which provides evidence
that the shape of the pulse spectrum has an effect on the quenching
of the molecular formation rate~\cite{SALZMANN2005}.
We have shown here that phase-shaping of the applied fs pulses can
be used to control the quenching rate.
%
%
This result provides evidence that the quenching is a coherent
process, although we are unable to distinguish the specific
excitation pathway stimulated by the fs pulses. Further experimental
and theoretical study of fs pulses for optimal control of
photoassociative ultracold molecule formation is warranted,
particularly for the creation of stable singlet molecules whose
deeper $X^1\Sigma_g^+$ potential may be better suited to the broad
inherent bandwidth of fs pulses.

\begin{acknowledgments}
We thank
Fran\c{c}oise Masnou-Seeuws, Christiane Koch,
Thorsten K\"{o}hler, and Nick Bigelow for helpful discussions. This
work was supported by the National Science Foundation, grant no.\
PHY9877023. AJD acknowledges support from an EPSRC DTA studentship.
\end{acknowledgments}

\end{document}